\documentclass[twocolumn,superscriptaddress,nofootinbib,aps,prl,floatfix,
preprintnumbers,amsmath,amssymb,groupedaddress]{revtex4-1}

\usepackage{dsfont}
\usepackage{epsfig}
\usepackage{slashed}
\usepackage{bbold}
\usepackage{psfrag}
\usepackage{color}
\PassOptionsToPackage{caption=false}{subfig}
\usepackage{subfig}
\usepackage{multirow}
\usepackage{booktabs}
\PassOptionsToPackage{hyphens}{url}
\usepackage{hyperref}
\usepackage{wasysym}

\begin{document}

\def\as{\alpha_S}
\def\t{{\bar t}}
\def\dy{{\Delta y}}
\def\dmody{{\vert\Delta y\vert}}
\def\Mtt{M_{t\bar t}}
\def\PT{P_{T,t\bar t}}
\def\GeV{\, \rm GeV}
\def\AFB{\rm A_{FB}}
\def\AFBincNUM{0.095 \pm 0.007}
\title{
  Top Quark Pair Production in Association with a Jet \\ with
  NLO QCD Off-Shell
  Effects at the Large Hadron Collider}

\author{\textsc{G. Bevilacqua$^{\,a}$, H. B. Hartanto$^{\,b}$, 
M. Kraus$^{\,b}$ and M. Worek$^{\,b}$} }
\affiliation{\vspace{.3cm}
  $^a$
INFN, Laboratori Nazionali di Frascati, Via E. Fermi 40, I-00044
Frascati, Italy  \\
$^b$
Institut f\"ur Theoretische Teilchenphysik und Kosmologie,
RWTH Aachen University, D-52056 Aachen, Germany}

\preprint{INFN-15-07/LNF, TTK-15-22}

\begin{abstract}
 We present a complete description of top quark pair production in
 association with a jet in the dilepton channel. Our calculation is
 accurate to next-to-leading order in QCD (NLO) and includes all
 non-resonant diagrams, interferences and off-shell effects  of the
 top quark.  Moreover, non-resonant and off-shell effects due to the
 finite W gauge boson width are taken into account.  This calculation
 constitutes the first fully realistic NLO computation for top quark
 pair production with a final state jet in hadronic collisions.
 Numerical results for differential distributions as well as total
 cross sections are presented for the Large Hadron Collider (LHC) at 8
 TeV.  With our inclusive cuts, NLO predictions reduce the unphysical
 scale dependence by more than a factor of $3$  and lower the total
 rate by about $13\%$  compared to leading order QCD (LO) predictions.
 In addition, the size of the top quark off-shell effects is estimated
 to be below  $2\%$.  
\end{abstract}
\maketitle

{\bf Introduction:}  Top quark studies are currently driven by the LHC
experiments. An exploration of  top quark production and  decay
dynamics is among the main physics goals of ATLAS and CMS. Besides the
determination of the top quark mass, key measurements at the LHC
include the total cross section,  kinematic distributions,  spin
correlations and top quark  couplings  to the $W$ and $Z$ bosons,
photon  and the Standard Model (SM) Higgs boson. Searches for rare top
quark decays to probe physics beyond the SM  also play a prominent
role in research  programs of both experimental collaborations. The
top quark, however,  is an extremely short lived resonance and only
its decay products can be detected experimentally. In general, for
comparison with data, theoretical predictions must include top quark
decays.  In the SM, a top quark decays almost exclusively to a $W$
boson and a $b$ quark.  The experimentally  cleanest top quark decay
channel comprises leptonic $W$ gauge boson decays.  The signature for
this channel consists of two well isolated and oppositely charged
leptons with high transverse momenta, $p_T$, large missing $p_T$ from
invisible neutrinos and  two jets, which originate from $b$
quarks. Due to the large collision energy at the LHC, $t\bar{t}$ pairs
are abundantly produced with large $p_T$, hence, the probability for
the top quark to radiate gluons is enough  to make the $t\bar{t}j$
final state measurable with high statistics. In fact, for $p_{Tj}>40$
GeV, about  half of the $t\bar{t}$ events are expected to be
accompanied  by an additional hard jet.  The correct description of
$t\bar{t}j$ production is, therefore, essential to study the top quark
pair production at the LHC. For example, $t\bar{t}j$  can be  employed
in the measurement of the top quark mass by studying normalized
differential distribution cross section with respect to its invariant
mass  \cite{Alioli:2013mxa}. Moreover, $t\bar{t}j$ constitutes an
important background to processes with multijet final states. The most
prominent being  the SM Higgs boson production in the vector boson
fusion   with the following decay chain $H\to W^+W^-  \to
\ell^+\nu_\ell\ell^-\bar{\nu}_\ell $
\cite{Rainwater:1999sd,Kauer:2000hi}. The $t\bar{t}j$ production
plays a very important role in searches for physics beyond the
SM. For example,   it is one of the main backgrounds to processes such
as supersymmetric particle production.  Here, depending on the specific
model, typical signals also include jets, charged leptons, and missing
$p_T$ due to the escaping lightest supersymmetric particle
\cite{Mangano:2008ha}. Anomalous production of additional jets
accompanying a $t\bar{t}$ pair could also be a sign of new physics
beyond the SM \cite{Gresham:2011dg}. 

The NLO corrections to $t\bar{t}j$ with stable top quarks have been
first calculated  in
\cite{Dittmaier:2007wz,Dittmaier:2008uj}. Afterwards, LO top quark
decays  in the narrow width approximation (NWA)  have been included
\cite{Melnikov:2010iu}. Subsequently, NLO top quark decays in the NWA,
including $t\to W b j$, have been added consistently
\cite{Melnikov:2011qx}.  A different path has been taken in
\cite{Kardos:2011qa,Alioli:2011as,Czakon:2015cla}, where matching to
parton shower programs has been worked out.  In  this case, however,
NLO corrections to  the $t\bar{t}j$ production  with stable top quarks
have only been taken into account, while  top quark decays, if
included, have been modeled within  parton shower frameworks.  
%
\begin{figure*}[t!]
\begin{center}
\includegraphics[width=1.0\textwidth]{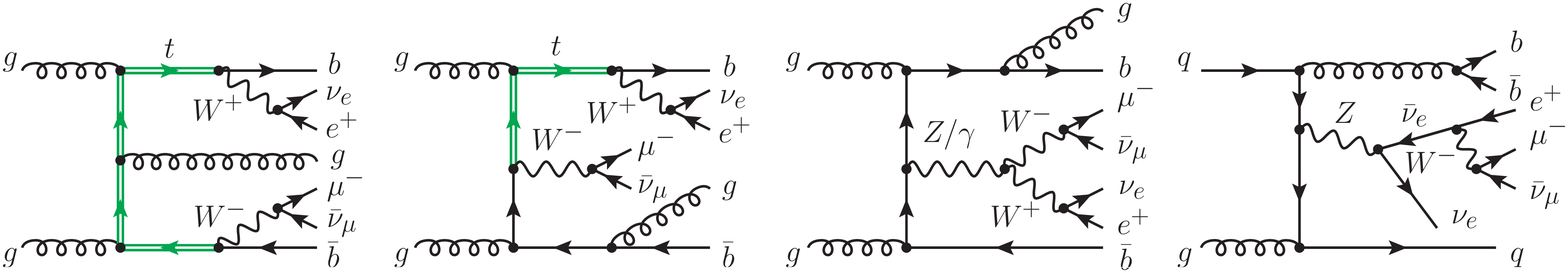}
\end{center}
\caption{\it Representative Feynman diagrams, involving two (first
  diagram), one (second diagram) and no top quark resonances (third
  diagram),  contributing to the leading order  $pp \to e^+\nu_e \mu^-
  \bar{\nu}_\mu b\bar{b}j$ process at ${\cal O}(\alpha_s^3 \alpha^4)
  $. The last diagram with a single $W$ boson resonance contributes to
  the off-shell  effects of the $W$ gauge boson.  }
\label{fig:fd}
\end{figure*}
%

In this Letter, we present a different approach. We drop altogether
the approximation that top quarks are only produced on-shell, and
concentrate on the fully realistic final state $pp \to e^+\nu_e \mu^-
\bar{\nu}_\mu b\bar{b} j+X$. We consistently take into account
resonant and non-resonant top quark contributions and all interference
effects among them. In addition, non-resonant and off-shell effects
due to the finite $W$ gauge boson width are included. Due to their
insignificance we neglect flavor mixing as well as contributions from
the suppressed initial bottom quark contributions.  A few examples of
Feynman diagrams contributing to the leading order process at ${\cal
O}(\alpha_s^3\alpha^4)$ are presented in Figure
\ref{fig:fd}.  We stress here 
that contributions of the order $O(\alpha_s
\alpha^6)$   have not been included in our
calculations. Full off-shell top quark effects at NLO have already
been considered in the literature for a simpler process, i.e.  top
quark pair production, first in
\cite{Denner:2010jp,Bevilacqua:2010qb}, and subsequently in
\cite{Denner:2012yc,Frederix:2013gra,Cascioli:2013wga,Heinrich:2013qaa}.
Quite recently, a first attempt to go beyond the NWA for a $2\to 5$
processes has been undertaken in \cite{Denner:2015yca}, where NLO
corrections to $pp \to e^+\nu_e \mu^- \bar{\nu}_\mu b\bar{b} H$ have
been considered.
 
{\bf Calculation:}  NLO QCD corrections to $pp \to e^+\nu_e \mu^-
\bar{\nu}_\mu b\bar{b} j$ have been calculated with the
\textsc{Helac-Nlo} Monte Carlo program \cite{Bevilacqua:2011xh}. This
is the first such computation with five final states (the decay
products of the $W$'s are not counted, because they do not couple to
color charged states) carried out within this framework.  We compute
the virtual corrections in the 't Hooft-Veltman version of the
dimensional  regularisation using
\textsc{Helac-1Loop} \cite{vanHameren:2009dr} and \textsc{CutTools}
\cite{Ossola:2007ax}, which are  based on the
Ossola-Papadopoulos-Pittau (OPP) reduction technique
\cite{Ossola:2006us,Ossola:2008xq,Draggiotis:2009yb}. The most
complicated one-loop diagrams in our calculations are heptagons. A
number of optimizations have been devised in the algorithm of
\textsc{Helac-1Loop} for the selection of loop topologies, which
discard in advance all  possibilities that are not compatible with
the SM. This allowed us to substantially  reduce the
generation time.  The process under consideration requires a special
treatment of unstable top quarks, which is achieved within the complex
mass scheme \cite{Denner:1999gp,Denner:2005fg}.  At the one loop level
the appearance of a non-zero top quark width in the propagator
requires the evaluation of scalar integrals with complex masses, which
is supported by the \textsc{OneLOop} program,  used  for the
evaluation of the  integrals \cite{vanHameren:2010cp}. For
consistency, mass renormalization for the top quark is also done by
applying the complex  mass scheme in the well known on-shell mass
counter term. The preservation of gauge symmetries by this approach is
explicitly checked by studying  Ward identities up to the one loop
level. Reweighting techniques, helicity and color sampling methods are
additionally used in order to optimize  the performance of our system.
The singularities from soft or  collinear parton emissions are
isolated via subtraction methods for NLO QCD
calculations. Specifically, two independent  subtraction schemes are
employed: the commonly used Catani-Seymour dipole subtraction
\cite{Catani:1996vz,Catani:2002hc,Czakon:2009ss}, and a fairly new
Nagy-Soper subtraction scheme \cite{Bevilacqua:2013iha}, both
implemented  in the \textsc{Helac-Dipoles} software
\cite{Czakon:2009ss}.  The implementation consists of a phase space
integrator of subtracted real radiation and  integrated subtraction
terms for  massless and massive cases. The phase space integration is
performed with the multichannel Monte Carlo generator \textsc{Phegas}
\cite{Papadopoulos:2000tt}  and \textsc{Kaleu}
\cite{vanHameren:2010gg}. In the latter case, dedicated additional
channels  for each subtraction term have been added for both
subtraction schemes to improve the convergence of the phase space
integrals for the subtracted  real contribution.  Let us also note,
that we have implemented a new option in \textsc{Helac-Nlo} for
automatically selecting the desired perturbative order in $\alpha_s$
and $\alpha$, preserving at the same time the structure and the
advantages of the Dyson-Schwinger recursive approach for the
construction of the amplitude.  This modification is particularly
useful in the current project, since we are interested in mixed
contributions,  i.e. ${\cal O}(\alpha^3_s\alpha^4)$ at LO and  ${\cal
  O}(\alpha^4_s\alpha^4)$ at NLO.

\begin{figure}[t!]
\begin{center}
\includegraphics[width=0.46\textwidth]{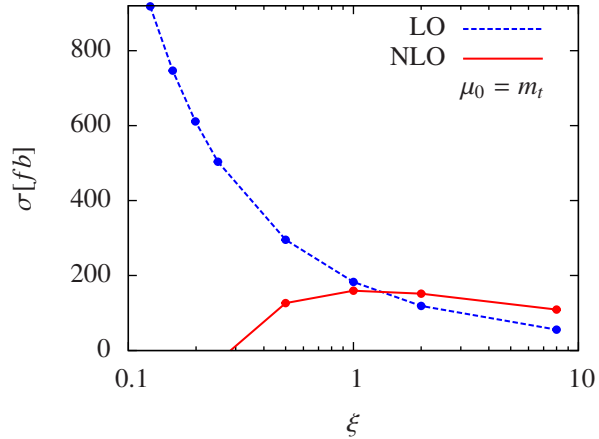}
\end{center}
\caption{\it Scale dependence of the LO and NLO cross sections for the
  $pp\to e^+\nu_e \mu^- \bar{\nu}_\mu b\bar{b} j +X$  process at the
  LHC for	$\sqrt{s}=8 ~{\rm TeV}$. The scale is set to a common
  value $\mu_{\rm R}=\mu_{\rm F}=\xi \mu_0$, where $\mu_0 = m_{\rm
    t}$.}
\label{fig:scale-variation}
\end{figure}

{\bf Phenomenological Application:} In the following we present our
numerical results for  $pp \to e^+\nu_e \mu^-\bar{\nu}_\mu
b\bar{b}j+X$ at the LHC  at the  center-of-mass energy of $\sqrt{s}=8
~{\rm TeV}$.  Decays of the weak bosons to different lepton
generations  are considered, to avoid virtual photon singularities
arising from $\gamma \to \ell^+ \ell^-$ decays. These effects are at
the level of $0.5\%$, as checked by an explicit LO calculation.  The
SM parameters are set to 
\begin{equation*}
\begin{array}{ll}
 G_{\rm F}=1.16637 \cdot 10^{-5} ~{\rm GeV}^{-2}\,,  
&   m_{\rm t}=173.3 ~{\rm GeV}\,,
\vspace{0.2cm}\\
 m_{\rm W}=80.399 ~{\rm GeV}\,, &\Gamma_{\rm W} = 2.09974 ~{\rm
                                  GeV}\,, 
\vspace{0.2cm}\\
 m_{\rm Z}=91.1876  ~{\rm GeV}\,, &\Gamma_{\rm Z} = 2.50966 ~{\rm GeV}\,, 
\end{array}
\end{equation*}
\begin{equation*}
\begin{array}{ll}
 \Gamma_{\rm t}^{\rm LO} = 1.48132 ~{\rm GeV}\,, &
 \Gamma_{\rm t}^{\rm NLO} = 1.3542 ~{\rm GeV}\,.
\end{array}
\end{equation*}
The top quark width has been calculated according to \cite{Jezabek:1988iv}. 
We use the MSTW2008  set of parton distribution
functions (PDFs)\cite{Martin:2009iq} , i.e.
MSTW2008lo68cl PDFs with a 1-loop running $\alpha_s$  at LO and
MSTW2008nlo68cl  with a 2-loop running $\alpha_s$ at NLO. 
All light quarks including $b$ quarks, as well as leptons, are treated as
massless. The suppressed contribution from $b$ quarks in the
initial state is neglected.  When considering the total
cross section at LO this contribution amounts only to $0.8\%$ of the
total cross section. The renormalization and factorization scale is
set to a common value $\mu_{\rm R}=\mu_{\rm F}=\mu_0=m_{\rm t}$.  Let
us notice  that  while evaluating  $\Gamma^{\rm NLO}_{\rm t}$ the
value of $\alpha_s$ at the scale $m_{\rm t}$  has been calculated from
$\alpha_s (m_{\rm Z})=0.118$. However, the  $\alpha_s(m_{\rm t})$ used
within \textsc{Helac-Nlo} is obtained from the NLO MSTW2008 set that
assumes $\alpha_s (m_{\rm Z})=0.12018$. As a  consequence, the
corresponding $\alpha_s (m_{\rm t})$ has a slightly different value.
Should we use this value in the calculation of the top  quark width, we
would rather get $\Gamma_{\rm t}^{\rm NLO} = 1.35207 ~{\rm GeV}$.  The
difference with respect to our value is at the level of one permille
only, therefore, completely negligible for  our NLO QCD results.  Jets
are defined out of partons with pseudorapidity $|\eta|<5$  by the
anti-$k_{\rm T}$ jet algorithm \cite{Cacciari:2008gp} with the
separation parameter $R=0.5$. We require exactly two $b$-jets, at least
one light jet, two charged leptons and missing $p_T$.   Final
states have to fulfill the following kinematical requirements 
\begin{equation*}
\begin{array}{lcl}
 p_{T \ell}>30 ~{\rm GeV}\,,    & & p_{Tj}>40 ~{\rm GeV}\,, 
\vspace{0.2cm}\\
p^{\rm miss}_{T} >40 ~{\rm GeV} \,,   & \quad \quad \quad 
& \Delta R_{jj}>0.5\,,
\vspace{0.2cm}\\
\Delta R_{\ell\ell}>0.4 \,,  && 
 \Delta R_{\ell j}>0.4 \,,
\vspace{0.2cm}\\
 |y_\ell|<2.5\,,&&
|y_j|<2.5 \,, 
\end{array}
\end{equation*}
where $\ell$ stands for  $\mu^-,e^+$ and $j$ corresponds to light- and
$b$-jets. Results for the total cross sections are
\begin{equation}
\begin{array}{lll}
\sigma^{\rm LO}_{\rm \textsc{Helac-Nlo}} 
&=& 183.1^{\, +112.2 \, (61\%)}_{\,\,\,\,  -64.2 \, (35\%)} ~{\rm fb}\,,
\vspace{0.2cm}\\
\sigma^{\rm NLO}_{\rm \textsc{Helac-Nlo}} 
&=& 159.7^{\, -33.1 \,(21\%)}_{\,\,\,\, -7.9 \, (\,\,\, 5\%)} ~ {\rm fb}\,.
\end{array}
\end{equation}
At the central scale, the full $pp$ cross section receives  negative
and moderate NLO  corrections of $13\%$.  Theoretical uncertainties,
associated with neglected  higher order terms in the perturbative
expansion, have been estimated by varying the renormalization and
factorization scales in $\alpha_s$ and PDFs, up and down by a factor
of 2 around the central scale of the process, i.e. $\mu_0$.  The scale
uncertainties are evaluated to be $61\%$ ($48\%$ after symmetrization)
at LO and $21\%$ ($13\%$ after symmetrization) at NLO. Thus, a
reduction  of the theoretical error by a factor of 3 is observed. The
graphical presentation of the behavior of  LO and NLO cross
sections  upon varying the scale by a factor of $\xi$ with $\xi \in
\left\{ 0.125, \dots, 8\right\}$ is shown in Figure
\ref{fig:scale-variation}.  
%
\begin{figure}
\begin{center}
\includegraphics[width=0.46\textwidth]{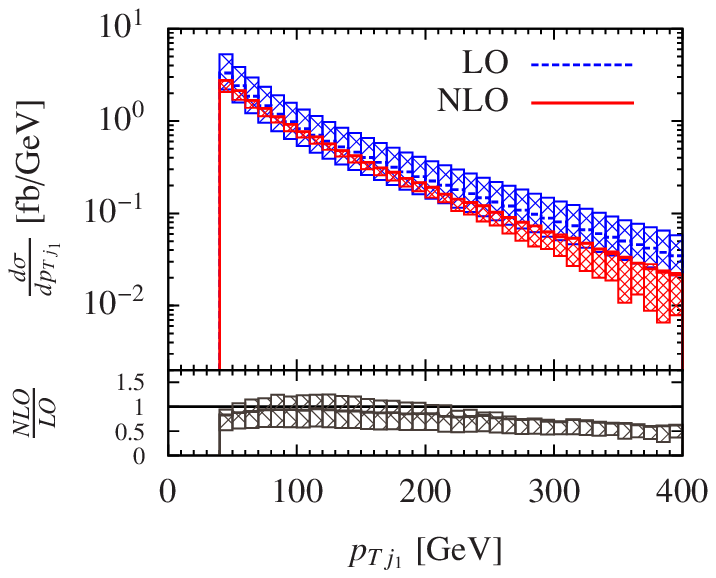}
\vspace{0.5cm}\\
\includegraphics[width=0.46\textwidth]{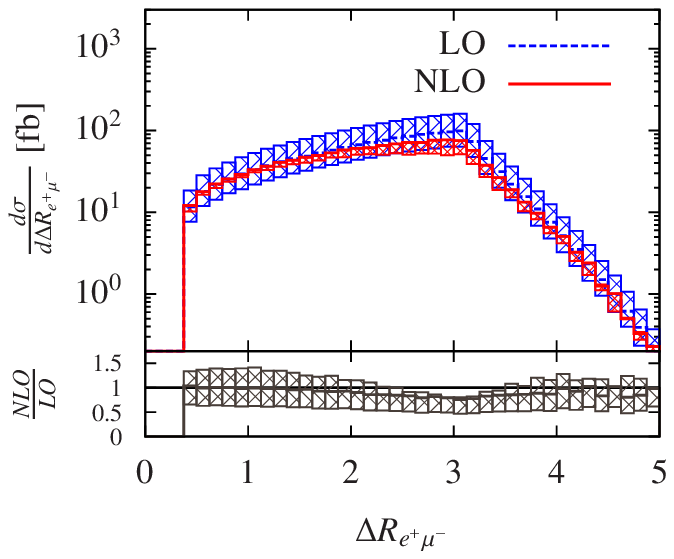}
\vspace{0.5cm}\\
\includegraphics[width=0.46\textwidth]{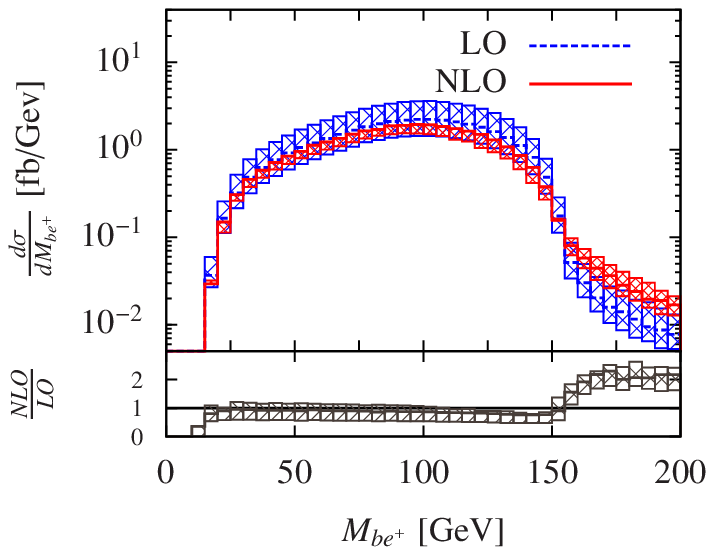}
\end{center}
\caption{\it   Transverse  momentum of the hardest light jet,  $\Delta
  R_{e^+\mu^-}$ and minimal $M_{be^+}$ for $pp\to e^+\nu_e \mu^-
  \bar{\nu}_\mu b\bar{b} j +X$  at the LHC with $\sqrt{s} = 8 ~{\rm
    TeV}$.  The uncertainty bands depict the scale variation. The
  lower panel displays the differential K factor and its uncertainty
  band.}
\label{fig:distributions}
\end{figure}
%

In the next step, the size of the non-factorizable corrections for our
setup is assessed. To that end, the full result has been compared with
the result in the NWA, which has been obtained by rescaling the $t\to
W b$ coupling and $\Gamma_{\rm t}$ by a small factor to mimic the
limit  $\Gamma_{\rm t} \to 0$.  Finite top quark width effects change
the cross section by less than $1\% \,(2\%)$ at LO (NLO), which is
consistent with the expected uncertainty of the NWA, i.e. of the order
of ${\cal O}(\Gamma_{\rm t}/m_{\rm t})$.  We have also calculated the
NLO cross section with a setup  from Ref. \cite{Melnikov:2011qx},
where NLO QCD corrections in the NWA have been evaluated for the $pp
\to e^+ \nu_e e^- \bar{\nu}_e b\bar{b}j$ final state. Instead of using
the top quark width from \cite{Melnikov:2011qx} we have calculated it
afresh to account for the off-shell effects of the $W$ gauge boson and
obtained $\Gamma_{\rm t}=1.31844$ GeV. In addition,  we have included
bottom quark contributions in the initial state and required at least
two $b$-jets in the final state. Our finding is  $\sigma^{\rm
  NLO}_{\rm \textsc{Helac-Nlo}}= (275.5 \pm 0.6) ~{\rm
  fb}$. Comparing
to the result from  \cite{Melnikov:2011qx} we observe a $4.5\%$
difference, which is of the order of the NWA accuracy for the top
quark and the $W$ gauge boson. However, further investigation of the
sources of the discrepancy would be desirable. We leave it for future
study.

Representative differential distributions are presented in Figure
\ref{fig:distributions}. We exhibit the transverse momentum of the
hardest (in $p_T$) light jet, $p_{Tj_1}$,  the  separation between
charged leptons in the rapidity azimuthal angle plane,  $\Delta
R_{e^+\mu^-}$, and the minimal invariant mass of the positron and
$b$-jet, $M_{be^+}$.  The dashed (blue) curve corresponds to the LO,
whereas the solid (red) one to the NLO result. The upper panels show
the distributions themselves and the scale dependence bands that  are
constructed by calculating, bin-by-bin, a maximal and a minimal value
out of  the following set $\left\{m_{\rm t}/2,m_{\rm t},2m_{\rm t}\right\}$.
The lower  panels display the differential ${\cal K}$ factor.
Higher order corrections to $p_{Tj_1}$  do not simply rescale the
shape of the LO distribution, but instead induce distortions.   With a
fixed scale choice  they reach $-50\%$ within the plotted range. Thus,
the $p_{Tj_1}$ differential cross section can only be properly
described when the higher order corrections are taken into account.
Therefore, LO calculations together  with a suitably chosen global
${\cal K}$ factor would not approximate the full NLO QCD calculation
well enough. However, a nearly  constant ${\cal K}$ factor can be
achieved with a judicious choice of the dynamic scale. Negative NLO
corrections in the high $p_T$  tail  means that the LO result is
higher than the NLO one.  The dynamic scale should depend on the $p_T$
of the hardest jet, and its value should increase  in the tail of 
the distribution. On the other hand, the asymptotic freedom guarantees
that the value of $\alpha_s$ becomes smaller  there, resulting in
lower NLO and LO cross sections.  Because of the different dependence
on the scale (see Figure \ref{fig:scale-variation}), the LO cross
section, which in general is much more sensitive to the variation of
the scale, will change more rapidly than the NLO curve,  driving a
positive NLO/LO ratio in this region.  We leave the search for such a
scale for the future.   On the contrary,  for the $\Delta
R_{e^+\mu^-}$ distribution, negative, moderate and quite stable
corrections are visible. This  can be explained by the dimensionless
nature of the observable. Certainly, $d\sigma/d\Delta R_{e^+\mu^-}$
receives contributions from all scales, most notably from those that
are sensitive  to the threshold for the $t\bar{t}j$
production. Indeed, for our scale choice, effects of the phase space
regions close to this  threshold dominate and a dynamic scale will
not alter this behavior.  Finally, the invariant mass distribution of
the positron and $b$-jet is shown.  In general, one cannot determine,
which $b$-jet  should be paired with the positron. To increase the
probability that both final states come  from the decay cascade
initiated by the same top quark we select the  $be^+$ pair, that
returns the smallest invariant mass \cite{Beneke:2000hk}. In case of
the  $t\bar{t}$ production  this observable has proved to be
particularly important for extracting $m_{\rm t}$  with a very high
precision \cite{Chatrchyan:2013boa,Aad:2015nba}. The top quark mass
can be determined either from the shape of the distribution away from
the kinematical end-point, defined as $M_{be^+} =\sqrt{m^2_{\rm  t}
  -m^2_{\rm W}} \approx 153.5 ~{\rm GeV}$, or from the behavior of the
observable in the vicinity of that point. In the former case,
off-shell effects are negligible, in the latter  they might even reach
$50\%$ \cite{AlcarazMaestre:2012vp}. When the  top quark and $W$ gauge
boson decay on-shell, the end-point is represented by a sharp cut.
However, additional radiation and off-shell effects introduce a
smearing to the region, which is highly sensitive to the details of the
description of the process. Thus, off-shell effects might  prove to be
very important for $t\bar{t}j$ as well, should  top quark mass
measurements be carried out using $M_{be^+}$.

{\bf Summary and Outlook:} In this Letter, NLO QCD corrections to $pp
\to e^+\nu_e\mu^-\bar{\nu}_\mu  b\bar{b}j+X$ with complete off-shell
and interference effects have been calculated for the first time.  We
have shown that NLO QCD corrections to the total cross section are
moderate ($13\%$). Nevertheless, their impact    on some differential
distributions is much larger. We have presented two cases,
$p_{Tj_{1}}$ and $M_{be^+}$, where higher order corrections are
indispensable to correctly describe the whole range of the
observable. We have also estimated the size of the top quark off-shell
effects at  NLO for the total cross section, and confirmed that they
are of the order of  ${\cal O}(\Gamma_{\rm t}/m_{\rm t})$. On the
other hand, their influence on differential distributions might be
much stronger, as has already been suggested by studies for the
$pp\to e^+\nu_e \mu^- \bar{\nu}_\mu b\bar{b}$  production process
\cite{AlcarazMaestre:2012vp}. We leave further comparisons for the
future. 

{\bf Acknowledgments:} 
We would like to thank A. van Hameren for providing us with a new
version of the \textsc{Kaleu} software and C. G. Papadopoulos for
useful discussions.  The work of H. B.  Hartanto was supported by the
German Research Foundation (DFG) under Grant {\it "Subtraction schemes at
next-to-next-to-leading order with applications to top-quark and jet
physics"}. M. Worek and M. Kraus acknowledge support by the DFG under
Grant No. WO 1900/1-1 $-$ {\it ``Signals and Backgrounds Beyond
Leading Order. Phenomenological studies for the LHC''}.

\end{document}